\documentclass[prd,preprint,12pt]{article}
\usepackage[dvips]{graphicx}
\usepackage{endnotes}
\usepackage{amssymb}
\usepackage{amsmath}
\usepackage{graphicx}
\usepackage{multicol}

\begin{document}
\title{The interacting generalized Ricci dark energy model in non-flat universe}
\author{Masashi Suwa$^{}$\footnote{email: suwa.masashi@nihon-u.ac.jp}  , Koji Kobayashi and Hisashi Oshima}
\date{\it \small $^1$School of Pharmacy, Nihon University, 7-7-1, Narashinodai, Funabashi-shi, Chiba, 274-8555, Japan}
\maketitle
\begin{abstract}
We extend our previous analysis and consider the interacting holographic Ricci dark energy (IRDE) model in non-flat universe. 
We study astrophysical constraints on this model using the recent observations including the type Ia supernovae (SNIa), 
the baryon acoustic oscillation (BAO), the cosmic microwave background (CMB) anisotropy, and the Hubble parameter. 
It is shown that the allowed parameter range for the fractional energy density of the curvature 
is $-0.005$ $\lesssim$ $\Omega_{k0}$ $\lesssim$ $0.015$ in the presence of the interactions between dark energy and matter. 
Without the interaction, the flat universe is observationally disfavored in this model.

\end{abstract}


\section{Introduction}
The current astrophysical  observations of the Type Ia supernovae (SNIa)~\cite{riess1}, the cosmic microwave background (CMB)~\cite{WMAP}  and the  large scale structure (LSS) ~\cite{SDSS} have revealed that the expansion of our universe is accelerated~\cite{riess1}. This indicates that there exists  some unknown energy, called dark energy,  to realize the accelerated expansion.
The simplest interpretation of dark energy is the cosmological constant. 
However, this model  requires  an incredible fine-tuning, since the observed cosmological constant is extremely small compared to the fundamental Planck scale $\rho_{\Lambda} \sim 10^{-120}M_{p}^4$. Also, this model suffers from the cosmic coincidence problem: why the cosmological constant and matter have comparable energy density today even though their time evolution is so different.

Among various attempts to solve these problems, we focus on  the holographic dark energy (HDE) models~\cite{li}  motivated by the holographic principle of quantum gravity~\cite{quantumgravity}.  
Requiring that the total vacuum energy of  a system with size $L$ would not exceed the mass of the black hole of the same size,  the dark energy density is proposed as 
 \begin{eqnarray}
   \rho_{\Lambda} = 3c^2 M_{p}^2 L^{-2},
 \end{eqnarray}
where $c$ is a dimensionless parameter, $M_P = 1/ \sqrt{8 \pi G}$ is the reduced Planck mass.  
As for the size $L$, which is regarded as an IR cutoff, various possibilities are 
discussed in literatures, such as the Hubble parameter $L^{-1} =H$ ~\cite{li}, 
the future event horizon  $L =  R_h$ ~\cite{li},
the age of our universe  $L=T$ ~\cite{age}, and 
the Ricci scalar curvature  $L^{-2} = \cal{R}$ ~\cite{gao}. 
In our previous work~\cite{suwa}, we studied the Ricci dark energy (RDE) model with $L^{-2} = \cal{R}$ by introducing an interaction between dark energy and matter. It was shown that a nonvanishing interaction rate $Q \propto H \rho_\Lambda$ is favored by the observations~\cite{suwa}. 

In this paper, we extend our previous analysis, and consider the interacting RDE (IRDE) model  
in non-flat universe. 
This paper is organized as follows. 
In Sec. \ref{IRDE}, we describe the generalized IRDE model in non-flat universe,
and obtain analytic expressions for cosmic time evolution.
In Sec. \ref{obs}, we discuss the observational constraints on this model. 
We summarize our results in Sec. \ref{conc}.


\section{The interacting Ricci Dark Energy model}
\label{IRDE}
We study the interaction Ricci Dark Energy (IRDE) model in non-flat universe. 
The Friedmann-Robertson-Walker metric non-flat univrerse is given by 
\begin{eqnarray}
\label{metric}
 ds^{2}= dt^{2} - a^{2}(t)
                \left(
                         \frac{dr^{2}}{1-kr^{2}} +r^{2}d \theta^{2} 
                                   +r^{2}\sin^{2}{\theta} d \phi^{2}
                \right),
  \end{eqnarray}
where $k=1$, 0, $-1$ for closed, flat, and open geometries. 
The Friedmann equation in non-flat univrerse takes the form
 \begin{eqnarray}
  H^{2} = \frac{1}{3M_{p}^{2}} 
                (\rho_{\Lambda}+\rho_{m}+\rho_{\gamma}+\rho_{k}),
  \label{Friedmann}
  \end{eqnarray}
where $\rho_{\Lambda}, \rho_{m}, \rho_{\gamma}$ and $\rho_{k}$ represent energy density of dark energy, 
matter, radiation and curvature, respectively, and $H=\dot{a}/a$ is the Hubble parameter.

We generalize the energy density of the Ricci dark energy as
 \begin{eqnarray}
 \label{rde}
  \rho_{\Lambda}
       &=& 3 M_{p}^2 \left(\frac{\alpha}{2} \partial_{x} H^{2} 
                            + \beta H^2 +\epsilon k e^{-2x}\right),
                \label{RDE}
 \end{eqnarray}
 where $\alpha$, $\beta$ and $\epsilon$ are dimensionless parameters 
 and $x=\ln{a}$. 
 In the case of $\beta=2 \alpha, \epsilon=\alpha, k=0$, this model is reduced to the ordinary RDE model \cite{suwa}. 
Moreover, we introduce a phenomenological interaction  between dark energy and matter. 
The energy densities $\rho_{\Lambda}$ and $\rho_{m}$ obey the following equations~\cite{IHDE}
 \begin{eqnarray}
 \label{CLL}
  \dot{\rho_{\Lambda}} + 3H (1+w_{\Lambda})\rho_{\Lambda} &=& -Q,\\
  \label{CLM}
    \dot{\rho_{m}} + 3H \rho_{m} &=& Q.
   \end{eqnarray}
We adopt the interaction rate given by
 \begin{eqnarray}
  \label{intrate}
   Q_{1}=\gamma H \rho_{\Lambda},
 \end{eqnarray}
  where $\gamma$ is a dimensionless parameter~\cite{suwa}.
To solve $H$, combining with eqs. (\ref{RDE}) and (\ref{CLM}), the Friedmann equation (\ref{Friedmann})
 is transformed as 
  \begin{eqnarray}
\nonumber
    \frac{ \alpha}{2} \frac{d^{2}H^{2}}{dx^{2}} 
    - \left(1-\frac{3\alpha}{2} -\frac{\alpha \gamma}{2} -\beta \right)\frac{d H^{2}}{dx} 
    -(3-3\beta-\beta \gamma) H^{2}  
    &&\\
\label{EOH}    
                -\Omega_{\gamma 0} H_{0}^{2} e^{-4x}
                -\{1- \epsilon(1+\gamma ) \}  k e^{-2x}
   &=&0.
 \end{eqnarray}
The solution to eq. (\ref{EOH}) is given by 
 \begin{eqnarray}
 \label{hubbleparameter}
  \frac{H^{2}}{H_{0}^{2}}    &=& A_{+} e^{\sigma_{+} x} + A_{-} e^{\sigma_{-} x}
            + A_{\gamma} e^{-4 x} + A_{k} e^{-2 x} ,
\label{solution}
 \end{eqnarray}
where 
 \begin{eqnarray}
 \label{sigmapm}
  \sigma_{\pm} 
   &=& \frac{(2-3\alpha - \alpha \gamma-2\beta)
          \pm \sqrt{(2+3\alpha-2\beta)^{2}
                     +\alpha\gamma(6\alpha +\alpha \gamma -4\beta -4)   }  }
         {2\alpha} ,
 \end{eqnarray}
   $\Omega_{\gamma 0}=\rho_{ \gamma 0}/\rho_{c0}$,  
 $\Omega_{k 0}= - k/H_{0}^{2}$ and  
 $\rho_{c0}=3M_{p}^{2}H_{0}^{2}$. 
 When $\sigma_{\pm}$ can be imaginary for sufficiently large $\alpha$, $\beta$ and $\gamma$, 
 $H^{2}$ has oscillatory behavior~\cite{suwa}.
 The constants $A_{\gamma}$, $A_{k}$ and $A_{\pm}$ are obtained as
 \begin{eqnarray}
  A_{\gamma} = \frac{\Omega_{\gamma 0}}{(2 \alpha - \beta) (1- \gamma)+1 } ,
 \end{eqnarray}
 \begin{eqnarray}
  A_{k} = \frac{1-(1+ \gamma)\epsilon }{1+( \alpha - \beta) (1+ \gamma) }\Omega_{k 0} ,
 \end{eqnarray}
 \begin{eqnarray}
 \label{rholambda}
 A_{\pm} &=& \pm \frac{ \alpha(\sigma_{\mp}+4) A_{\gamma}
            +\alpha(\sigma_{\mp}+2) A_{k}
                + 2 \Omega_{\Lambda 0}+2\epsilon\Omega_{k0}
                -\left(\alpha \sigma_{-} +2\beta\right) }
               {\alpha(\sigma_{+}-\sigma_{-})} .  
 \end{eqnarray}
In the case of $\beta=2 \alpha$, $\epsilon = \alpha$ and $\Omega_{k0}=0$, 
eq. (\ref{solution}) reduces to the result obtained in our previous work~\cite{suwa}.
 Substituting eq. (\ref{solution}) to eq. (\ref{RDE}), 
 the Ricci dark energy density is obtained as 
 \begin{eqnarray}
 \label{rdes}
  \rho_{\Lambda}
    &=&   \rho_{c0} 
         \left\{ 
              \sum_{i} 
              \left(\frac{ \alpha \sigma_{i}}{2}+\beta
                                    \right)
                                    A_{i} e^{\sigma_{i}x} 
              -\epsilon  \Omega_{k0} e^{-2x}
       \right\} ,
 \end{eqnarray}
 where $i\in\{+,-,\gamma,k\}$, $\sigma_{\gamma}=-4, \sigma_{k}=-2$,
 Likewise, the matter density is obtained as 
 \begin{eqnarray}
 \label{rms}
  \rho_{m}
  &=&   \rho_{c0}
         \left\{
                \sum_{i}\left(1 - \frac{\alpha  \sigma_{i}}{2} -\beta\right)A_{i} e^{\sigma_{i}x}
      - \Omega_{\gamma 0} e^{-4x} 
       -(1-\epsilon) \Omega_{k 0} e^{-2x}
       \right\} .
  \end{eqnarray}
 To derive the equation of state parameter $w_{\Lambda}$ of the Ricci dark energy,
 substituting eq. (\ref{rdes}) into the following expression:
 \begin{eqnarray}
  w_{\Lambda} 
    =  -1-\frac{1}{3}
             \left(
                      \gamma + \frac{1}{\rho_{\Lambda}}\frac{d\rho_{\Lambda}}{dx}
             \right).
  \end{eqnarray}


\section{Observational constraints}
\label{obs}
In this section, we discuss cosmological constraints on the IRDE model in the non-flat universe ($k\neq 0$) obtained from SNIa, CMB, BAO and the Hubble parameter observations.

The luminosity distance in the non-flat universe can be written as 
 \begin{eqnarray}
 d_{L} = \frac{(1+z)}{H_{0} \sqrt{\Omega_{k0}}} \sinh \left( H_{0} \sqrt{\Omega_{k0}} \int_{0}^{z} \frac{dz'}{H(z')} \right).
 \end{eqnarray}
The SNIa observations measure the distance modulus $\mu$ 
of a supernova and its redshift $z$. 
The distance modulus is given by 
 \begin{eqnarray}
  \mu = 5 \log_{10}{\frac{d_{L}}{\rm Mpc}}. 
 \end{eqnarray}
 We use the Union data set of 580 SNIa~\cite{580SN} to obtain limits on the relevant parameters $\alpha, \gamma$ and $\Omega_{\Lambda 0}$ 
 by minimizing $\chi_{\rm SN}^{2}$~\cite{chi2}.

The CMB shift parameter $R$ is given by  
 \begin{eqnarray}
  R &=& \sqrt{ \frac{\Omega_{m0}}{ \Omega_{k0}} }
                \sinh \left( H_{0} \sqrt{\Omega_{k0}} \int_{0}^{z_{\rm CMB}} \frac{dz}{H(z)}  \right), 
 \end{eqnarray}
 where $z_{\rm CMB}=1089$ is the redshift at recombination, 
 and $\Omega_{m0}=\rho_{m0}/\rho_{c0}$ is the matter fraction at present.
We use the value $R=1.725\pm0.018$ obtained 
 from the WMAP9 data~\cite{WMAP}. 
The CMB constraints are obtained by minimizing $\chi_{\rm CMB}^{2}$~\cite{chi2}. 
The shift parameter gives a complementary bound to the SNIa data ($z\lesssim 2$), 
since this parameter involves the large redshift behavior ($z\sim 1000$).

Signatures of the baryon acoustic oscillation (BAO) are provided by the observations of large-scale galaxy clustering. The BAO parameter $A$ is defined by
  \begin{eqnarray}
  A &=& \sqrt{\Omega_{m0}}
               \left(
                       \frac{H_{0}}{H(z_{\rm BAO})}
               \right)^{1/3}
               \left[
                       \frac{1}{z_{\rm BAO} \sqrt{\Omega_{k0}}} 
                       \sinh \left( H_{0} \sqrt{\Omega_{k0}} 
                                   \int_{0}^{z_{\rm BAO}} \frac{dz}{H(z)}  
                             \right)
               \right]^{2/3},
     \end{eqnarray}
where $z_{\rm BAO}=0.35$. 
We use the measurement of the BAO peak in the distribution of luminous red galaxies (LRGs) observed 
 in SDSS~\cite{SDSS}: 
 \begin{eqnarray}
  A &=& 0.469
                         \left(
                                  \frac{0.95}{0.98}
                         \right)^{-0.35}
               \pm 0.017. 
 \end{eqnarray}
The BAO constraints are obtained by minimizing $\chi_{\rm BAO}^{2}$~\cite{chi2}.

 The Hubble parameter constraints are given by minimizing
 \begin{eqnarray}
  \chi_{H}^{2} = \sum_{i}^{28} \left( \frac{H(z_{i})-H_{obs}(z_{i})}{\sigma_{H_{i}}}   \right)^{2},
 \end{eqnarray}
where $\sigma_{H_{i}}$ is the $1\sigma$ uncertainty of the observational $H(z)$ data~\cite{Hubble}.



\begin{figure}[htbp] 
 \begin{center}
  \begin{minipage}{1.0\textwidth}
   \begin{center}
    \begin{minipage}{0.45\textwidth}
     \begin{center}
      \includegraphics[width=0.95\textwidth]{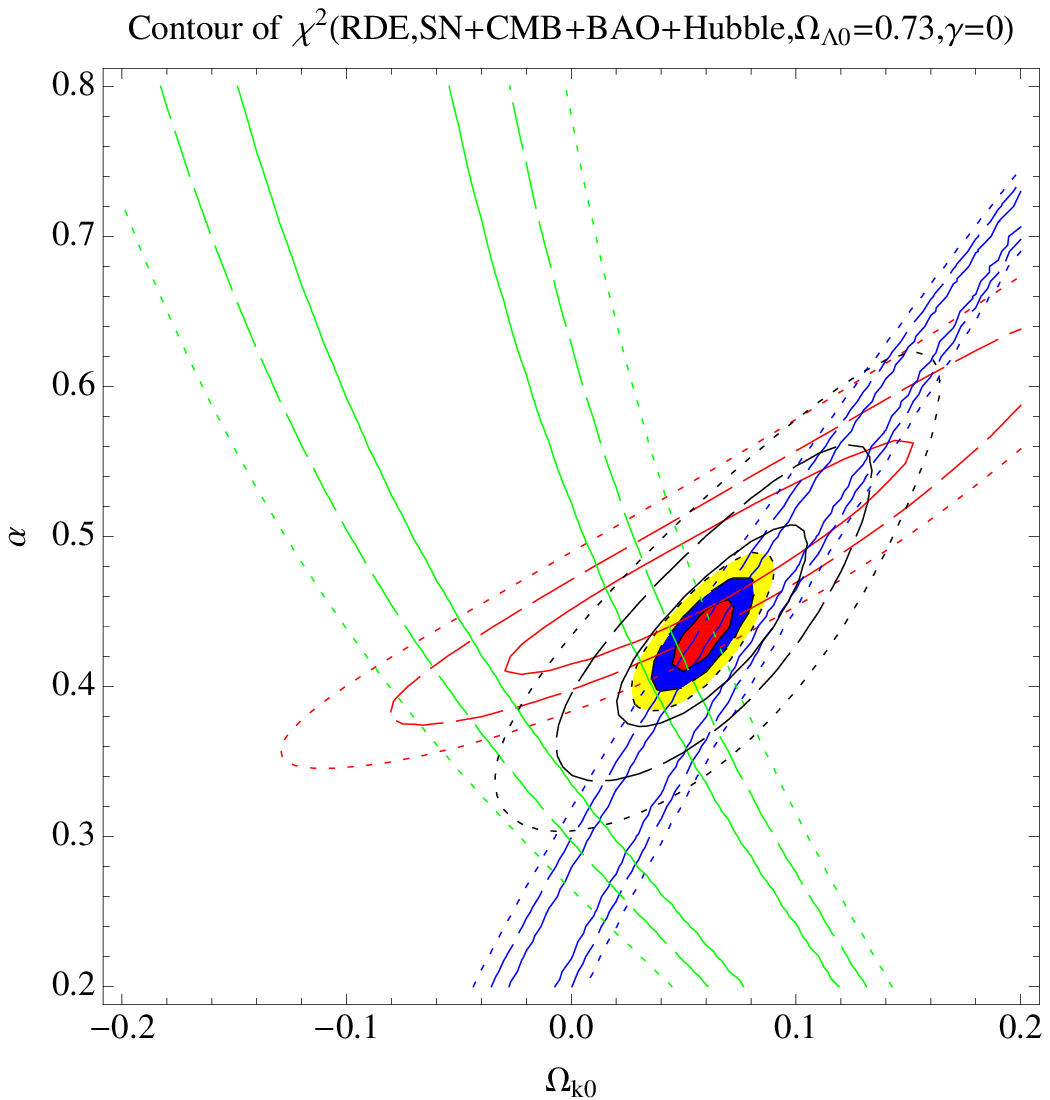}\\
        ~~~~(a) 
      \end{center}
     \end{minipage} 
     \begin{minipage}{0.45\textwidth}
      \begin{center} 
       \includegraphics[width=0.95\textwidth]{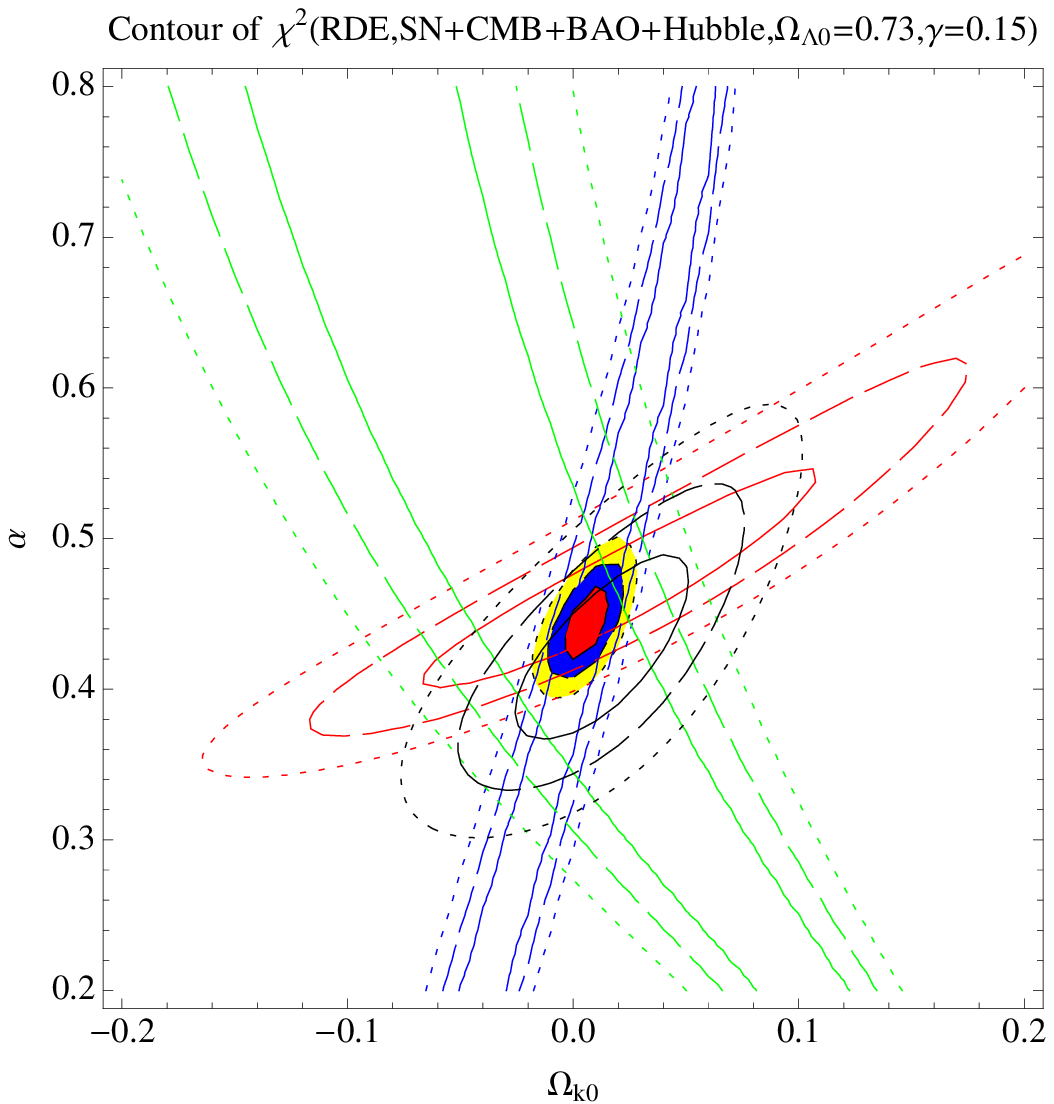}\\
        ~~~~ (b) 
      \end{center}
     \end{minipage}
    \end{center} 
   \end{minipage} 
\caption{
The probability contours for SNIa (red),
CMB (blue), BAO (green) and Hubble (black) observations in the ($\alpha, \Omega_{k0}$)-plane in the case without interactions ($\gamma=0$). 
The $1\sigma$, $2\sigma$ and $3\sigma$ contours are drawn with
solid, dashed and dotted lines, respectively.
The joined constraints using 
$\chi^{2}=\chi_{\rm SN}^{2}+\chi_{\rm CMB}^{2}+\chi_{\rm BAO}^{2}+\chi_{H}^{2}$ 
are shown as shaded contours. 
\label{fig:contours}
}
 \end{center}
\end{figure}

\begin{figure}[htbp]
\centering
\includegraphics[width=3.8in]
 {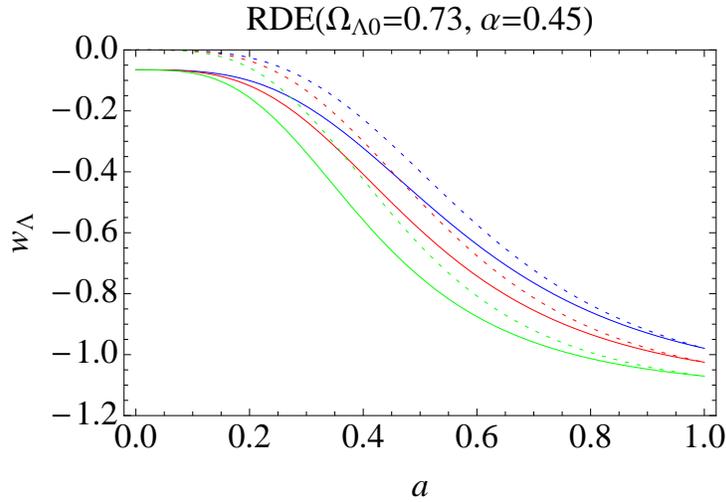}
\caption{
Plot of $w_{\Lambda}$ versus $a$ for RDE($\Omega_{\Lambda}=0.73, \alpha=0.45, \gamma=0$)
The lines for 
k=-0.1 (blue), 0 (red) and 0.1 (green)  
in the case without interactions (dotted line) and with interaction (solid line). 
}
\label{fig:w_a_gamma}
\end{figure}

In Fig.\ref{fig:w_a_gamma}, we plot the equation of state parameter $w_{\Lambda}$ for $\Omega_{k0}=-0.1$ (blue), 0 (red) 
and 0.1 (green) as a function of the scale factor $a$. 
The dotted and solid lines are the results for the case without interaction ($\gamma$=0) and with interaction ($\gamma=0.15$), respectively. 
The dark energy parameter is fixed as $\alpha=0.45$. 
Though the value $\Omega_{k0}= \pm 0.1$ is too large (see Fig. \ref{fig:contours}), 
we use these values to visualize the effect of the curvature in the figure. 
The condition for accelerated expansion at present $ w_{\Lambda}$ $<$ $-1/3$ is satisfied
for both cases.

\begin{figure}[p]
\centering
    \begin{minipage}{0.45\textwidth}
     \begin{center}
      \includegraphics[width=70.0mm]
       {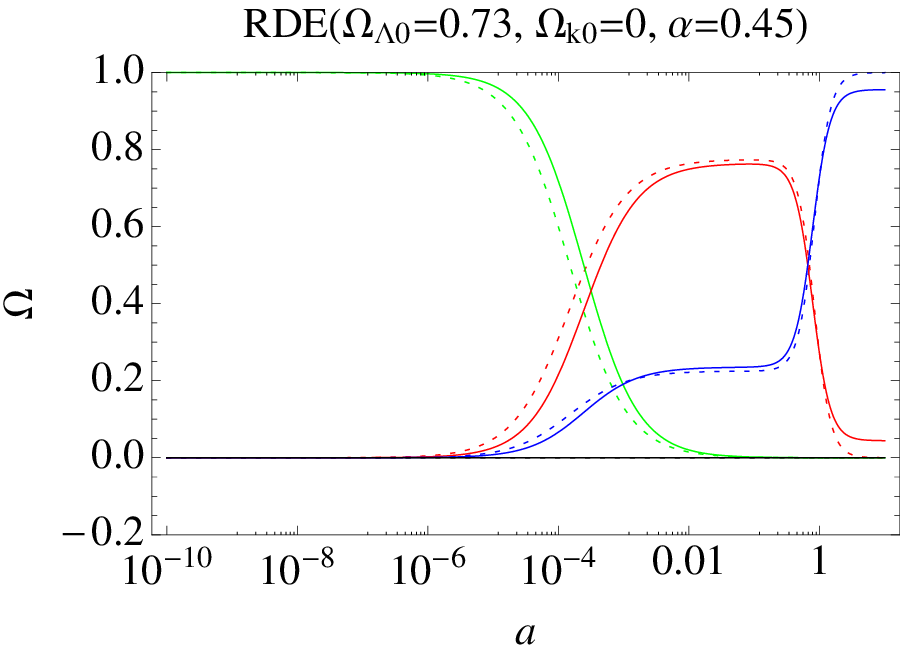}
      \end{center}
      \begin{center}
              ~~~~   (a) 
      \end{center}
     \end{minipage}
     \begin{minipage}{0.45\textwidth}
      \begin{center}
       \includegraphics[width=71.0mm]
        {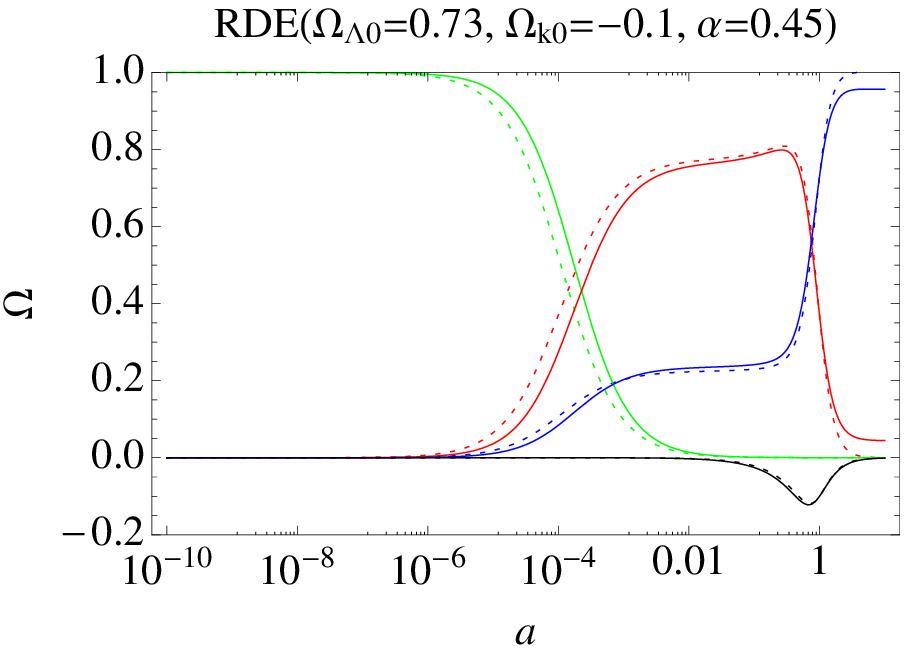}
      \begin{center}
              ~~~~~~   (b) 
      \end{center}
        \end{center}
       \end{minipage}
    \begin{minipage}{0.45\textwidth}
     \begin{center}
  \includegraphics[width=70.0mm]
   {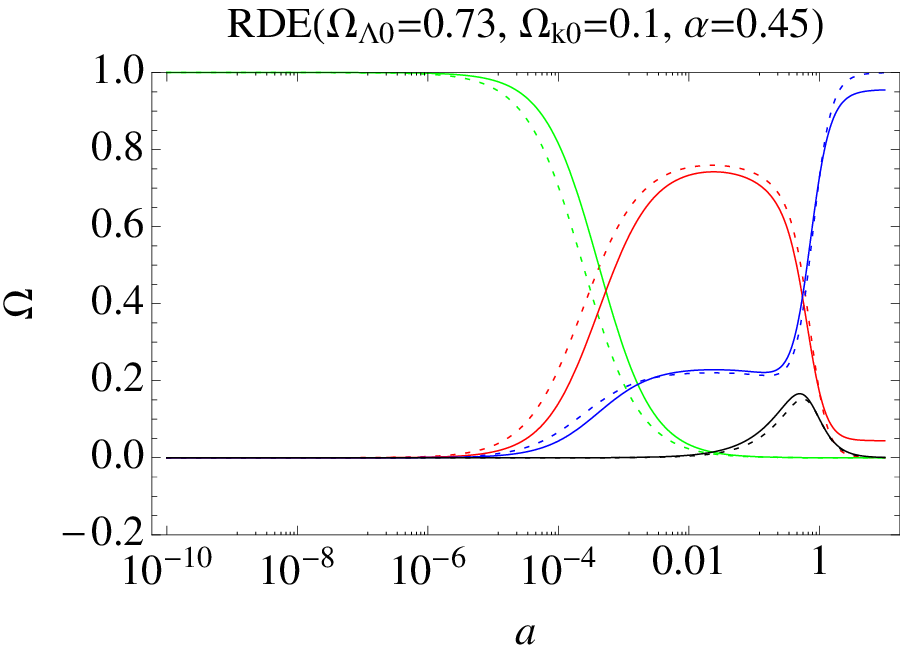}
      \begin{center}
              ~~~~   (c) 
      \end{center}
        \end{center}
       \end{minipage}

\caption{
Plot of $\Omega$ versus $a$ 
for RDE($\Omega_{\Lambda}=0.73, \alpha=0.45, \gamma=0.15$). 
The lines for 
Radiation (Green), matter (red), dark energy (blue) and curvature (black) 
in the case without interactions (dotted line) and with interaction (solid line). 
These figures describe in the case of $\Omega_{k0}=0$ (Fig. a), -0.1 (Fig. b) and 0.1 (Fig. c).
 \label{fig:Omega_a_gamma}
}
\end{figure}

In Fig.\ref{fig:Omega_a_gamma}, we plot the evolution of the energy density fractions 
$\Omega$ for radiation (green), matter (red) and dark energy (blue) for $\alpha=0.45$.
The dotted and solid lines are the results for the case without interaction ($\gamma$=0) and with interaction ($\gamma=0.15$), respectively. 
The panels (a), (b) and (c) corresponds to 
$\Omega_{k0}=0$, $-0.1$ and 0.1, respectively. 
One can see that the effect of the curvature can be important only for $a$ $\sim$ 1. 
The fractions $\Omega_{\Lambda}$ and $\Omega_{m}$ for $\Omega_{k0}=-0.1$ in the panel (b)  ($\Omega_{k0}=0.1$ in the panel (c)) 
are slightly increased (decreased) around 0.1 $\lesssim$ $a$ $\lesssim$ 1 
compared to the flat case (a). 
Around radiation-matter equality, 
the radiation component $\Omega_{\gamma}$ is increased by several percents
compared to the corresponding result without interaction (dotted line), 
while the $\Omega_{m}$ is decreased due to the interaction.


\section{Conclusion}
\label{conc}
We have considered the IRDE model in the non-flat universe.
We have derived the analytic solutions for 
 the Hubble parameter~(\ref{hubbleparameter}), the dark energy density~(\ref{rdes}) and matter energy density~(\ref{rms}).
We have also studied astrophysical constraints on this model using the recent observations 
including SNIa, BAO, CMB anisotropy, and the Hubble parameter.
We have shown that the allowed parameter range for the fractional energy density of the curvature 
is $-0.005$ $\lesssim$ $\Omega_{k0}$ $\lesssim$ $0.015$ 
for $\gamma$ $=$ 0.15. 
The best fit values with $1\sigma$ error are $\Omega_{k 0}= 0.006 \pm 0.010$  
and $\alpha = 0.45 \pm 0.03$ with $\chi_{\rm min}^{2}=580$.
We have shown that the IRDE model with a small curvature is allowed by observational constraints.
Without the interaction, the flat universe is observationally disfavored in this model.


\section*{Acknowledgment}

We would like to thank T. Nihei for his valuable discussion, helpful advice and reading the manuscript.




\begin{thebibliography}{99}
\bibitem{riess1}
A. G. Riess et al., Astron. J. 116 (1998) 1009; 
S. Perlmutter et al., Astrophys. J. 517 (1999) 565.
\bibitem{WMAP}
C. L. Bennet et al., Astrophys. J. Suppl. 148 (2003) 1; 
D. N. Spergel et al., Astrophys. J. Suppl.  148 (2003) 175; 
D. N. Spergel et al., Astrophys. J. Suppl. 170 (2007) 377; 
L. Page et al., Astrophys. J. Suppl. 170 (2007) 335; 
G. Hinshaw et al., Astrophys. J. Suppl. 170 (2007) 263; 
E. Komatsu et al., Astrophys. J. Suppl. 180 (2009) 330 [arXiv:0803.0547]; 
E. Komatsu et al. [WMAP Collaboration], Astrophys. J.
Suppl. 192, 18 (2011); 
G. Hinshaw, D. Larson, E. Komatsu, D. N. Spergel,
C. L. Bennett, J. Dunkley, M. R. Nolta and M. Halpern
et al., arXiv:1212.5226 [astro-ph.CO].
\bibitem{SDSS}
D. J. Eisenstein et al., Astrophys. J. 633 (2005) 560 
[arXiv:astro-ph/0501171]. 

\bibitem{li}
M. Li, Phys. Lett. B 603 (2004) 1. 
\bibitem{quantumgravity}
G. 't Hooft, arXiv:gr-qc/9310026; 
L. Susskind, J. Math. Phys. 36 (1995) 6377 [arXiv:hep-th/0403052]; 
J. M. Maldacena, Adv. Theor. Math. Phys. 2 (1998) 231; 
W. Fischler, L. Susskind, arXiv:hep-th/9806039.
\bibitem{einstein}
  A.~Einstein,
  Sitzungsber.\ Preuss.\ Akad.\ Wiss.\ Berlin (Math.\ Phys.\ ) 1917 (1917) 142.
\bibitem{weinberg}
See e.g., S. Weinberg, astro-ph/0005265.
\bibitem{cohen}
A. Cohen, D. Kaplan, A. Nelson, Phys. Rev. Lett. 82 (1999) 4971 
[arXiv:hep-th/9803132].
\bibitem{hsu}
S. D. H. Hsu, Phys. Lett. B 594 (2004) 13 [arXiv:hep-th/0403052].
\bibitem{age}
R.G. Cai,
Phys. Lett. B 657, 228 (2007);
\bibitem{gao}
C. Gao, F. Wu, X. Chen, Y. G. Shen, Phys. Rev. D 79 (2009) 043511 
[arXiv:0712.1394].

\bibitem{RDE}
L. Xu, J. Lu, W. Li, Eur. Phys. J. C 64 (2009) 89 
[arXiv:astro-ph/0906.0210]; 
C. J. Feng, X. Z. Li, Phys. Lett. B 680 (2009) 355; 
C. J. Feng, X. Z. Li, Phys. Lett. B 679 (2009) 151 
[arXiv:hep-th/0904.2976]; 
S. B. Chen, J. L. Jing, arXiv:0904.2950; 
C. J. Feng, X. Zhang, Phys. Lett. B 680 (2009) 399 
[arXiv:gr-qc/0904.0045].


\bibitem{suwa}
M. Suwa, T. Nihei, 
Phys. Rev. D 81, 023519.

\bibitem{IHDE}
B. Wang, C. Y. Lin, E. Abdalla, Phys. Lett. B 637 (2006) 357
[hep-th/0509107];
M. S. Berger, H. Shojaei, Phys. Rev. D 74 (2006) 043530 
[astro-ph/0606408];  
M. R. Setare, Phys. Lett. B 642 (2006) 1
[hep-th/0609069]; 
M. R. Setare, JCAP 0701 (2007) 023 
[hep-th/0701242]; 
Q. Wu, Y. Gong, A. Wang, J. S. Alcaniz, 
Phys. Lett. B 659 (2008) 34  
[arXiv:astro-ph/0705.1006]; 
C. Feng, B. Wang, Y. Gong, R. K. Su, JCAP 0709 (2007) 005 
[arXiv:astro-ph/0706.4033]; 
K. Karwan, JCAP 0805 (2008) 011 
[arXiv:astro-ph/0801.1755]; 
R. Horvat, Phys. Rev. D 70 (2004) 087301; 
D. Pavon, W. Zimdahl, Phys. Lett. B 628 (2005) 206; 
B. Guberina, R. Horvat, H. Nikolic, JCAP 0701 (2007) 012; 
X. Zhang, F. Q. Wu, Phys. Rev. D 72 (2005) 043524; 
Z. Chang, F. Q. Wu, X. Zhang, Phys. Lett. B 633 (2006) 14; 
X. Zhang, F. Q. Wu, Phys. Rev. D76 (2007) 023502; 
X. Zhang, Int. J. Mod. Phys. D 14 (2005) 1597; 
M. R. Setare, J. Zhang, X. Zhang, JCAP 0703 (2007) 007; 
J. Zhang, X. Zhang, H. Liu, Phys. Lett. B659 (2008) 26; 
X. Zhang, Phys. Lett. B 648 (2007) 1; 
X. Zhang, Phys. Rev. D74 (2006) 103505; 
J. Zhang, X. Zhang, H. Liu, Phys. Lett. B651 (2007) 84; 
Y. Ma, X. Zhang, Phys. Lett. B 661 (2008) 239; 
J. Zhang, X. Zhang, H. Liu, Eur. Phys. J. C 52 (2007) 693. 
\bibitem{newinteraction}
M. Li, X. D. Li, S. Wang, Y. Wang, X. Zhang, arXiv:0910.3855. 

\bibitem{580SN}
M. Kowalski et al., Astrophys. J. 686 (2008) 749 
[arXiv:astro-ph/0804.4142];
N. Suzuki et al.,Astrphys. J. 746 (2012) 85
[arXiv:astro-ph/1105.3470].

\bibitem{chi2}
S. Nesseris, L. Perivolaropoulos, Phys. Rev. D 72  (2005)  123519  
[arXiv:astro-ph/0511040]; 
L. Perivolaropoulos, Phys. Rev. D 71 (2005) 063503 
[arXiv:astro-ph/0412308]; 
S. Nesseris, L. Perivolaropoulos, JCAP 0702 (2007) 025  
[arXiv:astro-ph/0612653]; 
E. D. Pietro, J. F. Claeskens, Mon. Not. Roy. Astron. Soc. 341 (2003) 1299 
[arXiv:astro-ph/0207332]. 

\bibitem{Hubble}
Simon, J., Verde,L., \& Jimenez, R., Phys. Rev. D 71 (2005) 123001;
Gazta\~{n}aga, E., Cabr\'{e}, A., \& Hui, L. 2009, MNRAS, 399,1663;
Stern, D., et al. 2010, JCAP 1002 (2010) 008;
Moresco, M., et al. 2012, J. Cosmology Astropart. Phys., 1208, 006;
Busca, N. G., et al., [arXiv:astro-ph/1211.2616];
Omer Farooq, Data Mania, Bharat Ratra, [arXiv:astrp-ph/1308.0834].


\bibitem{cosmicconstraintsRDE}
L. Xu, W. Li, J. Lu, Mod. Phys. Lett. A 24 (2009) 1355 [arXiv:0810.4730]; 
X. Zhang, Phys. Rev. D 79 (2009) 103509 [arXiv:0901.2262]; 
C. J. Feng X. Z. Li, Phys. Lett. B 680 (2009) 184;
M. Li, X. D. Li, S. Wang, X. Zhang, JCAP 0906 (2009) 036 
[arXiv:astro-ph/0904.0928].

\bibitem{consistentobs}
M. Li, C. Lin, Y. Wang, JCAP 0805 (2008) 023;
Q. G. Huang, Y. G. Gong, JCAP 0408 (2004) 006 
[astro-ph/0403590]; 
X. Zhang, F. Q. Wu, Phys. Rev. D 72 (2005) 043524 
[astro-ph/0506310];
Z. Chang,  F. Q. Wu, X. Zhang, Phys. Lett. B 633 (2006) 14
[astro-ph/0509531];
X. Zhang,  F. Q. Wu, Phys. Rev. D 76 (2007) 023502 
[astro-ph/0701405]; 
H. Zhang, W. Zhong, Z. H. Zhu, S. He, Phys. Rev. D 76 (2007) 123508 
[arXiv:astro/0705.4409]; 
X. Chen, J. Liu, Y. Gong, Chin. Phys. Lett. 25 (2009) 3086 
[arXiv:gr-qc/0806.2415]. 
\bibitem{aspects}
E. Elizalde, S. Nojiri, S. D. Odintsov, P. Wang, Phys. Rev. D 71 (2005) 103504  
[hep-th/0502082]; 
S. Nojiri, S. D. Odintsov, Gen. Rel. Grav. 38 (2006) 1285 
[hep-th/0506212]; 
E. N. Saridakis, Phys. Lett. B 661 (2008) 335 [arXiv:gr-qc/0712.3806]; 
E. N. Saridakis, Phys. Lett. B 671 (2009) 331 [arXiv:gr-qc/0810.0645].


\bibitem{nonflat}
E. Ebrahimi, A. Sheykhi and H. Alavirad
arXiv:1209.3147;
A. Banijamali, M.R. Setare, B. Fazlpour
Int. J. Theor. Phys.  50 (2011) 3275-3283;
M.R. Setare
Phys. Lett.  B642 (2006) 1-4;
Q. G. Huang and M. Li
JCAP 0408 (2004) 013;
A. Sheykhi
Phys. Lett. B680 (2009) 113-117;
J. Zhang, M. Zhao, J. Lei and X. Zhang, 
Eur. Phys. J. C 74 (2014) 3178;
B. Majeed, M. Jamil, A. A. Siddiqui,
Int. J. Theor. Phys. 
DOI 10.1007/s10773-014-2197-3.


\bibitem{not_a_perfectly_flat}
K. Ichikawa et al JCAP 12 (2006) 005;
M. Jamil, M.U. Farooq, 
J. Cosmol. Astropart. Phys. 03, 001 (2010) ;
M. Jamil, E. N. Saridakis, 
J. Cosmol. Astropart. Phys. 7, 28 (2010) ;
M. Jamil, A. Sheykhi, 
Int. J. Theor. Phys. 50, 625 (2011);
M. Jamil, E. N. Saridakis, Setare, M.R.,
J. Cosmol. Astropart. Phys.11, 32 (2010a);
K. Karami, A. Sorouri, 
Phys. Scr. 82, 025901 (2010);
K. Karami,  et al., 
Gen. Relativ. Gravit. 43, 27 (2011a);
K. Karami,  et al., 
Europhys. Lett. 93, 69001 (2011b);
K. Karami, M. S. Khaledian, M. Jamil, 
Phys. Scr. 83, 025901 (2011c);
K. Karami, N. Sahraei, S. Ghaffari, 
J. High Energy Phys. 1108, 150 (2011d);
K. Karami, A. Sheykhi, N. Sahraei, S. Ghaffari, 
Europhys. Lett. 93, 29002 (2011e);
K. Karami, et al., Can. J. Phys. 90, 473 (2012a);
K. Karami, M. Jamil, M. Roos, S. Ghaffari, A. Abdolmaleki, 
Astrophys. Space Sci. 340, 175 (2012b).

\end{thebibliography}
\end{document}